\documentclass[12pt]{aa}
\usepackage[english]{babel}
\usepackage{psfig,graphicx}
\onecolumn

\begin{document}

\title{High--latitude supergiant V5112\,Sgr: enrichment of the envelope with heavy s-process metals}

\author{V.\,G.~Klochkova}

\date{\today}

\institute{Special Astrophysical Observatory,  Nizhnij Arkhyz, 369167, Russia}

\abstract{High--resolution (R\,=\,60\,000) echelle spectroscopy of the post--AGB supergiant V5112~Sgr
performed in 1996--2012 with the 6-meter telescope BTA has revealed peculiarities of the star optical
spectrum and has allowed the variability of the velocity field in the stellar atmosphere and envelope to be
studied in detail. An asymmetry and splitting of strong absorption lines with a low lower--level excitation
potential have been detected for the first time. The effect is maximal in Ba\,II lines whose profile is split
into three components. The profile shape and positions of the split lines change with time. The blue
components of the split absorption lines are shown to be formed in a structured circumstellar
envelope, suggesting an efficient dredge--up of the heavy metals produced during the preceding evolution
of this star into the envelope. The envelope expansion velocities have been estimated to be V$_{\rm exp} \approx$\,20 
and 30\,km/s. The mean radial velocity from diffuse bands in the spectrum of V5112\,Sgr coincides with
that from the short--wavelength shell component of the Na\,I~D lines, which leads to the conclusion about
their formation in the circumstellar envelope. Analysis of the set of radial velocities Vr based 
on symmetric absorption lines has confirmed the presence of pulsations in the stellar atmosphere with an amplitude
$\Delta Vr \le 8$\,km/s. 
\newline
{\it Keywords: stellar evolution, post-AGB stars, envelopes, spectra.}}

\titlerunning{\it  High--latitude supergiant V5112\,Sgr}
\authorrunning{\it Klochkova}

\maketitle

\section{Introduction}

The high-latitude supergiant V5112\,Sgr with Galactic coordinates l\,=\,$23\lefteqn{.}^{\rm o}98$, 
b\,=$-21\lefteqn{.}^{\rm o}0$   and spectral type  Sp\,=\,F2--F6 (Parthasarathy et al. 1988), identified with the 
strong infrared source IRAS\,19500$-$1709, belongs  to semiregular variable post-asymptotic giant branch 
(post--AGB) stars. Intermediate mass stars (their initial masses  are 3--8${\mathcal M}_{\odot}$) that evolve from the AGB, 
losing their matter through a stellar wind, are observed at  this short  stage. The secular variability of the main 
parameters observed in post--AGB stars stimulates their spectroscopic monitoring. A spectroscopic monitoring of selected 
AGB and post--AGB stars has been performed with the 6-meter  telescope BTA over the last decade. The main goal of the 
monitoring program is to reveal a peculiarity and probable variability of the spectrum and to study the temporal behavior 
of the velocity field in the extended atmosphere and envelope of peculiar supergiants. By now, based on observational 
data from the 6--m telescope, we have found spectroscopic variability in V510\,Pup (Klochkova and Chentsov 2004), 
BD+48$^{\rm o}$1220 (Klochkova et al. 2007a),  V2324\,Cyg (Klochkova et al. 2008a) and the optical counterparts of
the infrared sources IRAS\,01005+7910 (Klochkova et al. 2002) and IRAS\,20508+2011 (Klochkova et al. 2006).

Original results concerning the variability of the optical spectrum and the velocity field in the 
atmospheres have also been obtained for the variable stars CY\,CMi (Klochkova 1995; Klochkova
et al. 2007b), QY\,Sge (Klochkova et al. 2007c), V354\,Lac (Klochkova 2009; Klochkova et al. 2009),
and V448\,Lac (Klochkova et al. 2010). The BTA spectroscopy for post--AGB stars is presented in more
detail in our review (Klochkova 2012). Recently, Klochkova and Panchuk (2012) published the results
of their spectroscopic monitoring for the high--latitude supergiant LN\,Hya whose observations revealed a
peculiarity and variability of the profiles of strong Fe\,I, Fe\,II, Ba\,II, Si\,II, and other lines. 
Weak emissions of neutral atoms (VI, MnI, CoI, NiI, FeI) appeared in the June 1, 2010 spectrum. 
These features of the stellar spectrum detected for the first time suggest that the physical conditions 
in the upper atmospheric layers of LN\,Hya changed rapidly in 2010.

This paper is devoted to a detailed study of the optical spectrum for V5112\,Sgr. The study
of its photometric variability has a fairly long history (Arkhipova et al. 2010; Hrivnak et al. 2011).
However, as yet no definitive conclusions about the causes, type, and parameters of the photometric 
variability have been reached. The cyclic variability amplitude of V5112\,Sgr is low; it does not exceed 
$\rm \Delta V \le 0.2^m$ in V and changes with time. As regards the variability period, it has been determined 
incompletely: as Hrivnak et al. (2010) showed, the dominant period is 38$^{\rm d}$ in V.

A peculiarity of the infrared spectrum for IRAS\,19500$-$1709 is the presence of a peculiar
emission feature at 21\,$\mu$ (Kwok et al. 1999). The small group of objects with the as yet unidentified
21\,$\mu$ emission feature includes post--AGB stars with an atmospheric chemical composition changed
in the course of their own evolution. A generalization of the results of several publications appeared
in the late 1990's (Klochkova 1995; van Winckel et al. 1996; Reddy et al. 1997; van Winckel 1997)
led Klochkova (1997) and Decin et al. (1998) to conclude that the atmospheric chemical composition
of all post--AGB stars with the 21\,$\mu$ emission feature changed through the dredge-up of carbon
and heavy metals synthesized through the s--process.
Subsequent studies of an extended sample of objects confirmed this conclusion. At present, about a dozen
post--AGB stars with the 21\,$\mu$ emission feature have been studied in our Galaxy. Their main list is
contained in Hrivnak et al. (2010), a paper devoted to investigating the photometric variability of this type
of stars. 

IRAS\,19500$-$1709 is the highest-latitude object in the group of sources with the 21\,$\mu$  band.
The location of V5112\,Sgr at a high Galactic latitude already suggests that it belongs to an old population 
of the Galaxy. Its low metallicity (van Winckel et al. 1996) confirms this classification. The atmospheric 
chemical composition of the star studied in detail by these authors is typical of post--AGB stars
with the 21\,$\mu$ band: the atmospheres of these stars exhibit large overabundances of carbon and s--process
elements synthesized during their preceding evolutionary stages. As follows from Table~11 in van Winckel (1997),
V5112\,Sgr is the record--holder in the overabundances of carbon and s--process elements in the atmosphere.

The circumstellar envelope of V5112\,Sgr is also enriched with carbon and is a CO emission source
(van der Veen et al. 1993). The flux emitted by the circumstellar envelope in IRAS bands is comparable
to the visible flux from this star (Hrivnak et al. 1989).
Note that the infrared flux from most of the highlatitude F supergiants is low. In particular, no infrared
excess whatsoever was detected in the prototype star, the supergiant UU\,Her. The supergiant LN\,Hya,
the optical spectrum peculiarities and the velocity field variability in whose atmosphere were studied by
Klochkova and Panchuk (2012), is also a member of this small Galactic population (Sasselov 1984).

As can be seen from the above references, several papers devoted to investigating the chemical peculiarities 
of V5112\,Sgr have been published in the last 20 years, while there is virtually no information
about the peculiarities of its optical spectrum and its behavior with time. To make up for this deficiency, we
took new high--quality optical spectra for V5112\,Sgr in 1996--2012. Analysis of this observational material
provided previously unknown information concerning the variability and peculiarities of the star optical
spectrum and the detailed picture of radial velocities Vr in its atmosphere and envelope. In this paper, we
briefly describe the methods of spectroscopic observations and spectrum processing as well as present
and discuss our results.

\section{Observations, processing and analysis of spectra}

We obtained spectroscopic data for V5112\,Sgr at the Nasmyth focus of the 6-m BTA telescope
at the Special Astrophysical Observatory  with the NES echelle spectrograph designed by Panchuk et al. (2007,
2009). The observations were performed with a 2048$\times$2048--pixel CCD array and an image slicer
(Panchuk et al. 2009). The observations on August 2, 2012, were carried out with a larger 2048$\times$4096--pixel
CCD, which provided a considerable increase in the recorded spectral range. The spectral resolution is
$\lambda/\Delta\lambda \ge 60000$; the signal-to-noise ratio is S/N$\ge$100. The mean dates of observations (JD) and the
recorded spectral range are given in Table\,1.

One--dimensional spectra were extracted from two--dimensional echelle frames using the ECHELLE
context of the MIDAS software package modified by Yushkin and Klochkova (2005) by taking into
account the peculiarities of the NES optical scheme. Cosmic-ray hits were removed by a median averaging
of two spectra taken successively one after another. The wavelength calibration was made with a 
hollow--cathode Th–Ar lamp. The extracted spectroscopic data were processed with the DECH20t software
package developed by Galazutdinov (1992). Telluric [O I], O$_2$, and H$_2$O lines were used to check the
instrumental reconciliation of the spectra for the star and the hollow--cathode lamp. The procedure for
measuring the heliocentric radial velocity Vr from the NES spectra and the sources of errors are described 
in more detail in Klochkova et al. (2008b) and Klochkova and Tavolzhanskaya (2010). The
rms measurement error of Vr for stars with narrow  absorption lines in the spectrum is $\le$1.0 km/s (the
accuracy from a single line).

\begin{table}
\caption{Log of observations for V5112\,Sgr and measured heliocentric radial velocities Vr}
\begin{tabular}{ c c|  c| r   @{$\pm$} l| r  c|  r }
\hline
Date &JD & $\Delta\lambda,$ & \multicolumn{5}{c}{$V_{\rm r}$, km/s}\\ 
\cline{4-8}
     &2450000+& \AA{} &\multicolumn{2}{c|}{Metals}&\multicolumn{2}{c|}{NaI} &  DB \\
\hline     
  1         &  2    &    3       & \multicolumn{2}{c|}{4} & 5    &   6   & \hspace{0.5cm} 7 \\
\hline     
 05.07.1996&0269.43& 5150--8000 &+14.4&0.5(30)   &      &       & $-9.1$(4) \\  
 07.07.2001&2097.51& 4610--6074 &+21.1&0.1(190)  &+21.4 &$-9.6$ & $-10.5$(1) \\ 
 14.08.2006&3962.34& 4555--6007 & +5.4&0.1(237)  &+6.8  &$-8.5$ & $-9.5$(1)  \\ 
 28.09.2010&5462.26& 5165--6690 &+15.4&0.2(99)   &+9.4  &$-9.2$ &$-9.1$(2)   \\
 13.06.2011&5725.53& 4120--5580 &+8.5 &0.1(366)  &      &       & \\
 02.08.2012&6142.42& 3990--6980 &+10.1&0.1(377)  &+9.5  &$-8.8$ &$-8.8$ (5) \\  
\hline
\multicolumn{6}{c}{}&\multicolumn{2}{l}{Mean:} \\
\cline{7-8}
\multicolumn{6}{c}{} & $-9.0$ & $-9.1$ \\ 
\hline
\multicolumn{8}{l}{}\\ [-3mm]
\multicolumn{8}{l}{\footnotesize\it Note. Here, Vr (metals), Vr (NaI), and Vr (DB) are the mean velocities from}\\
\multicolumn{8}{l}{\footnotesize\it  symmetric absorptions, from the components of the Na I D lines, and from    }\\ 
\multicolumn{8}{l}{\footnotesize\it diffuse bands, respectively. The number of lines  used to determine the mean } \\ 
\multicolumn{8}{l}{\footnotesize\it velocities is given in parentheses.} \\
\end{tabular}
\label{Velocity}
\end{table}

\section{Main results}

In this section, we describe the main anomalies and variability of the line profiles as well as the related
peculiarities in the behavior of the velocity field in the atmosphere and envelope of V5112\,Sgr.

The overwhelming majority of absorption features in the stellar spectrum with low and moderate intensities have 
symmetric profiles without any anomalies. It is these numerous features that we used
to study the behavior of the star radial velocities with time. Having analyzed our sets of Vr based on
symmetric absorption lines, we concluded that there is no correlation between the line intensity and the
corresponding Vr. This allowed us to consider the averaged radial velocities for each spectrum in the
subsequent analysis. The mean velocity Vr(metals) reliably determined from numerous symmetric absorption lines 
is given in column~4 of Table~1. As follows from this table, Vr(metals) changes from date
to date with a small amplitude $\le 8$\,km/s relative to the systemic velocity Vsys\,=\,13\,km/s obtained by
Bujarrabal et al. (1992) and van der Veen (1993) from CO molecular lines. Obviously, this radial velocity
variability with a small amplitude is attributable to the stellar pulsations studied in detail by Hrivnak
et al. (2011). Based on a large set of measurements, these authors concluded that the radial velocity of
V5112 Sgr changes only slightly near its mean Vr$\approx$14\,km/s, which is very close to the systemic velocity
Vsys\,=\,13\,km/s.

{\bf The H$\alpha$ profile.} Emission components in the H$\alpha$ profile are typical of post-AGB stars (see examples
in Fig.~2 from Klochkova (1997)) and are among the main criteria for selecting such objects (Kwok 1993).
As follows from Fig.\,1, the H$\alpha$ profile contains emission components in all our spectra for V5112\,Sgr.
An H$\alpha$ profile with a double--peaked emission feature typical of post--AGB stars was recorded in 2010 and 2012. 
Having analyzed a large volume of spectroscopic observations for post--AGB stars near the H$\alpha$
line, Sanchez Contreras et al. (2008) classified the profiles of these stars. Using their classiﬁcation, we
attribute the H$\alpha$ profile in the spectrum of V5112\,Sgr in quiescent phases to the EFA (emission--filled absorption) type. 
According to universally accepted views, this profile type points to the existence of a long--lived circumstellar gas 
reservoir. The line width is determined primarily by the effect of scattering by free electrons and by the kinematics 
of the circumstellar structure.

\begin{figure}      
\includegraphics[angle=0,width=0.5\textwidth,bb=30 60 540 790,clip]{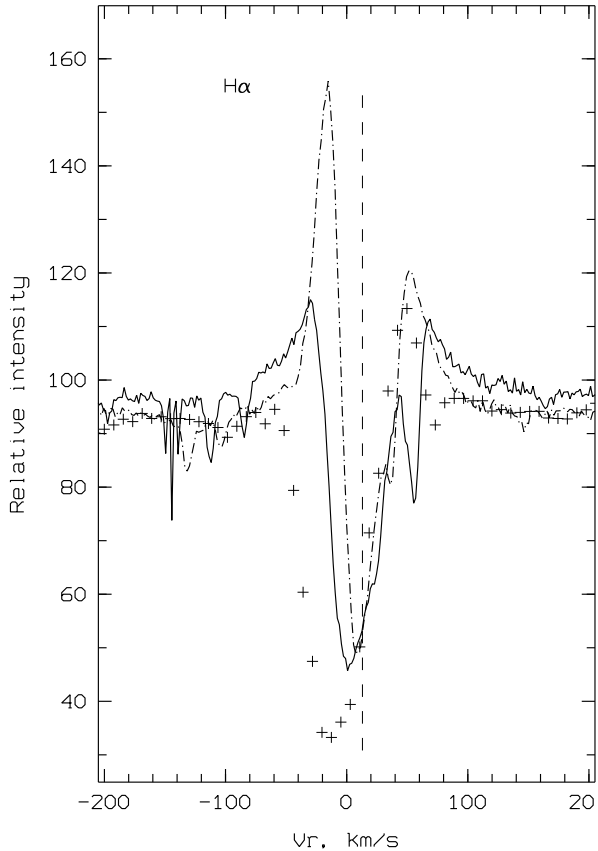}
\caption{H$\alpha$ profiles in the spectra of V5112\,Sgr taken at different dates from Table\,1: 
         August 2, 2012 (solid line) and September 28, 2010 (dash--dotted line). The crosses plot the profile 
         in the spectrum with resolution R\,=\,25000 taken on July 5, 1996, with the Lynx spectrograph (Panchuk et al. 1993). 
         Here and in subsequent figures, the vertical dashed line indicates the systemic velocity Vsys\,=\,13\,km/s.}
\end{figure}

Note the abrupt change of the H$\alpha$ profile type in the spectrum of V5112\,Sgr that occurred at the turn
of the new century: in 1996, we recorded a profile inverse to the P\,Cyg one, while in the 2000s, as follows
from Fig.\,1, the profile contained already two emission components. We emphasize that this change of the
H$\alpha$ profile type that we detected occurred concurrently with a significant change in the object  
photometric characteristics found by Hrivnak et al. (2010) during a long--term monitoring.

\begin{table}
\bigskip
\caption{Radial velocity Vr from three components of the Ba\,II lines in the spectra of V5112 Sgr taken at different dates}
\bigskip
\begin{tabular}{c|  c|   c|    r|   r   }          
\hline
\hspace{0.5cm} Date & \multicolumn{4}{c}{Vr, km/s}\\  
\cline{2-5}
               & \multicolumn{3}{c|}{BaII} & metals \\
\hline     
07.07.2001& $-17.3$& $-8.7$&+20.5  &  +21.07 \\ 
14.08.2006& $-14.9$& $-5.6$& +4.6  &  +5.37  \\ 
13.06.2011& $-16.4$& $-5.9$&+10.9  &  +8.48  \\
02.08.2012& $-17.0$& $-7.8$&+11.28 &  +10.12 \\   
\hline
\multicolumn{5}{l}{}\\ [-3mm]
\multicolumn{5}{l}{\footnotesize\it Note. The mean velocity inferred from the symmetric }\\ 
\multicolumn{5}{l}{\footnotesize\it   absorptions of metals from Table\,1 is given in the }\\
\multicolumn{5}{l}{\footnotesize\it   last column for comparison }\\    
\end{tabular}
\end{table}

The radial velocity derived from the absorption components of neutral hydrogen lines changes with
time (Fig.\,1). However, the small number of measurements (from one to four for different H\,I lines)
did not allow us to reveal any regular patterns in the behavior of Vr(H\,I).

{\bf Strong absorption lines.} Analyzing our highquality spectra of V5112\,Sgr taken in a wide wavelength range, 
apart from the variable H$\alpha$ profile, we revealed other, previously unknown, peculiarities of
the spectrum.

First, we found the strongest absorption lines of metals (Si\,II, Ba\,II, Y\,II, Zr\,II) originating in the star
high atmospheric layers to have anomalous profiles: asymmetric with an extended blue wing or split
into individual components. As an example, Fig.\,2 presents fragments of the spectra  with BaII\,4554 
and BaII\,4934\,\AA{} lines in the August 2, 2012 and July 7, 2001 spectra, respectively.
Anomalous split profiles of the resonance lines for various chemical elements are shown in Figs.\,3 and 4. 
A different width of the components is clearly seen here: for the long--wavelength component, 
it is approximately twice the width of the short--wavelength ones shifted considerably relative to the systemic velocity. 
This difference in widths suggests that the long--wavelength and blueshifted components are formed under different 
physical conditions. Note line  splitting can also be seen in that Ba\,II~4934\,\AA{}  the fragment of the spectrum 
for V5112\,Sgr from van Winckel (1997).

It should be emphasized that profile anomalies were detected only for the absorption lines of those
metals for which large overabundances were detected in the stellar atmosphere. At the same time, the
profiles of even the strongest absorption lines of ironpeak elements have no anomalies. This can be clearly
seen in Fig.\,2 from our comparison of the Ba\,II and Fe\,II lines with similar intensities.

\begin{figure}      
\includegraphics[angle=-90,width=0.5\textwidth,bb=30 50 540 780,clip]{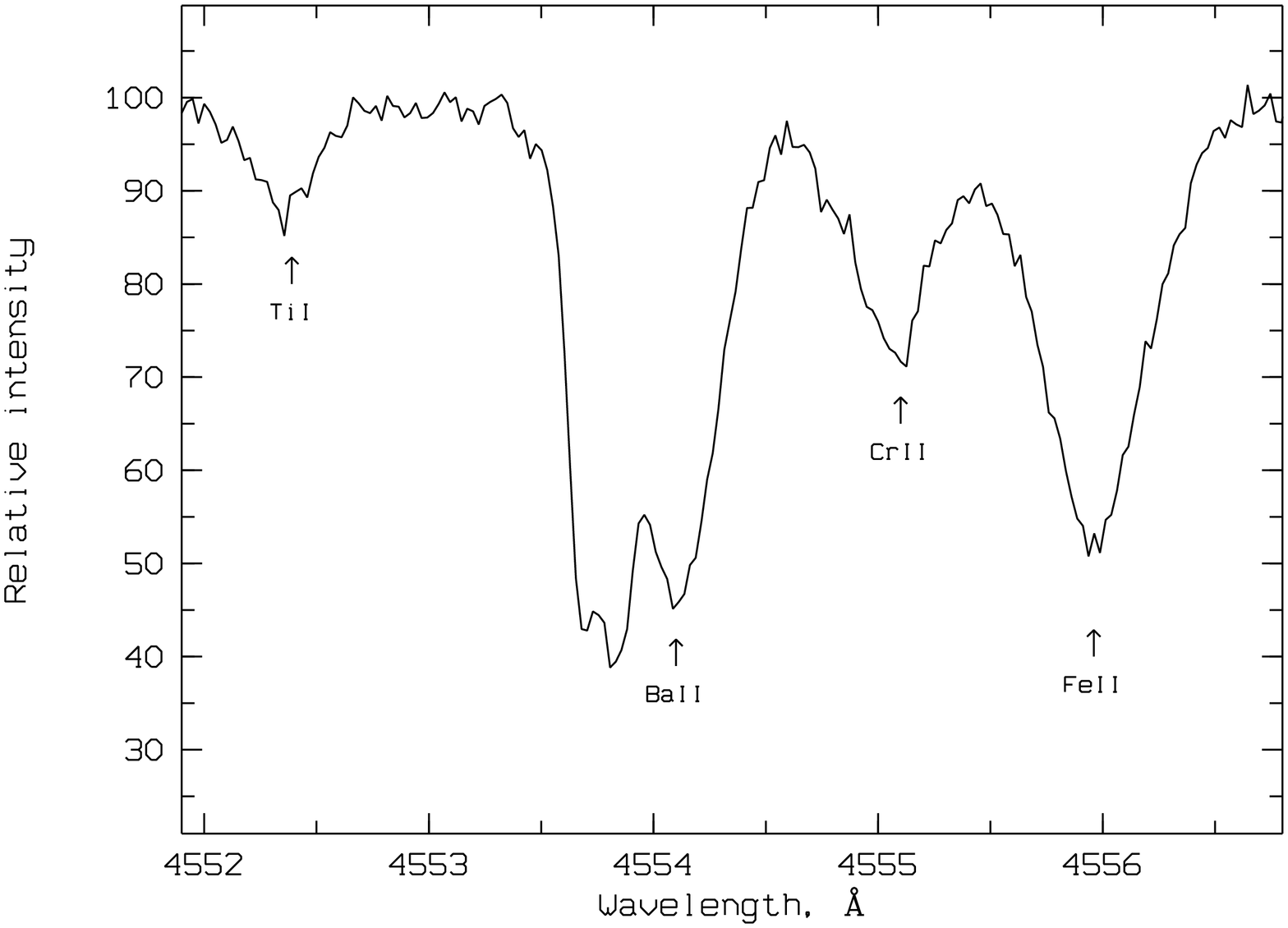}
\includegraphics[angle=-90,width=0.5\textwidth,bb=30 50 540 780,clip]{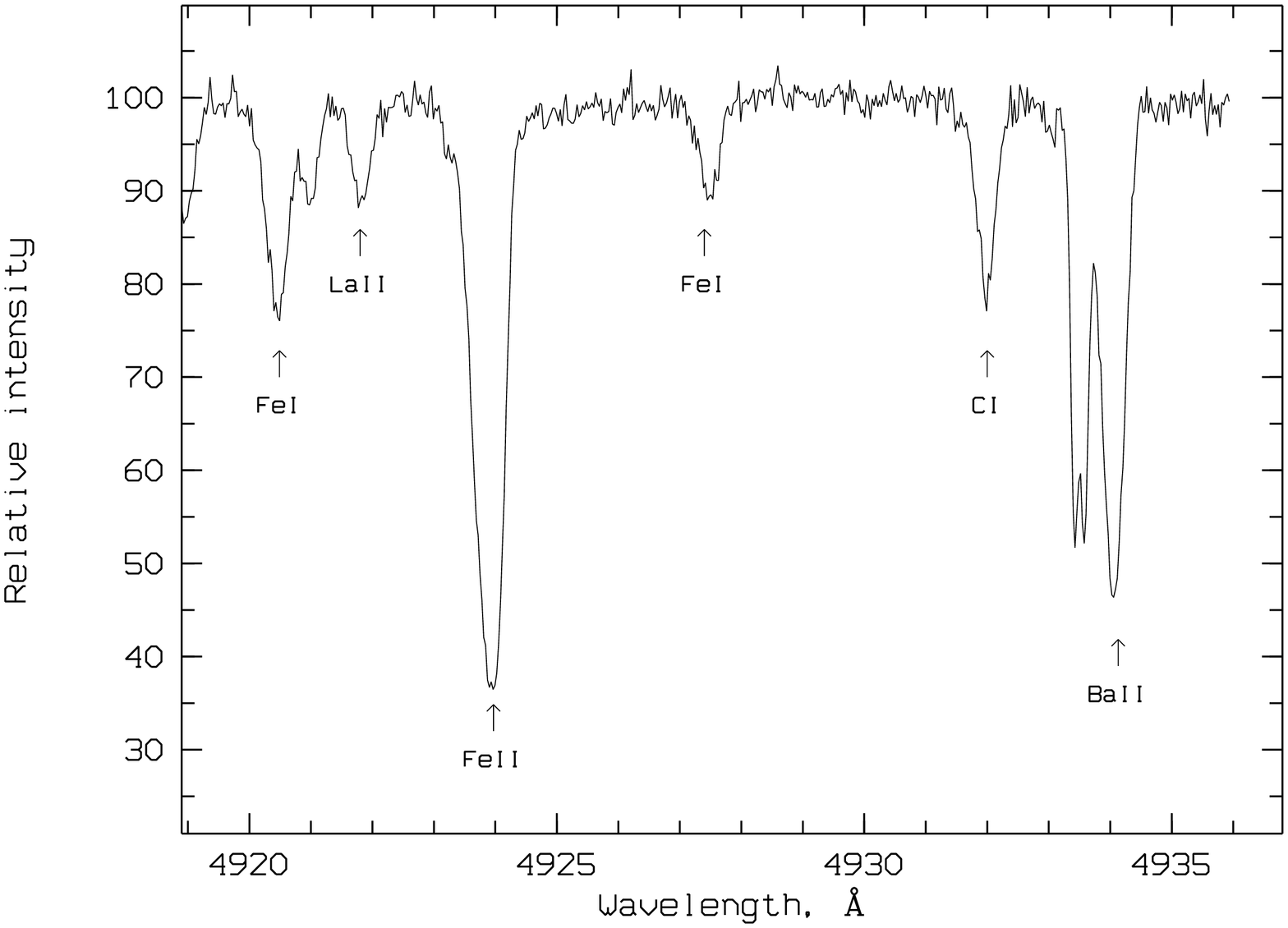}
\caption{Fragments of the spectra for V5112\,Sgr containing split lines. The Ba\,II\,4554\,\AA{}
       line in the  August 2, 2012 spectrum (left); the Ba\,II\,4934\,\AA{}  in the July 7, 2001 spectrum (right). 
       The identification of main absorption lines is indicated.}
\end{figure}

\begin{figure}  
\includegraphics[angle=0,width=0.4\textwidth,bb=30 50 540 780,clip]{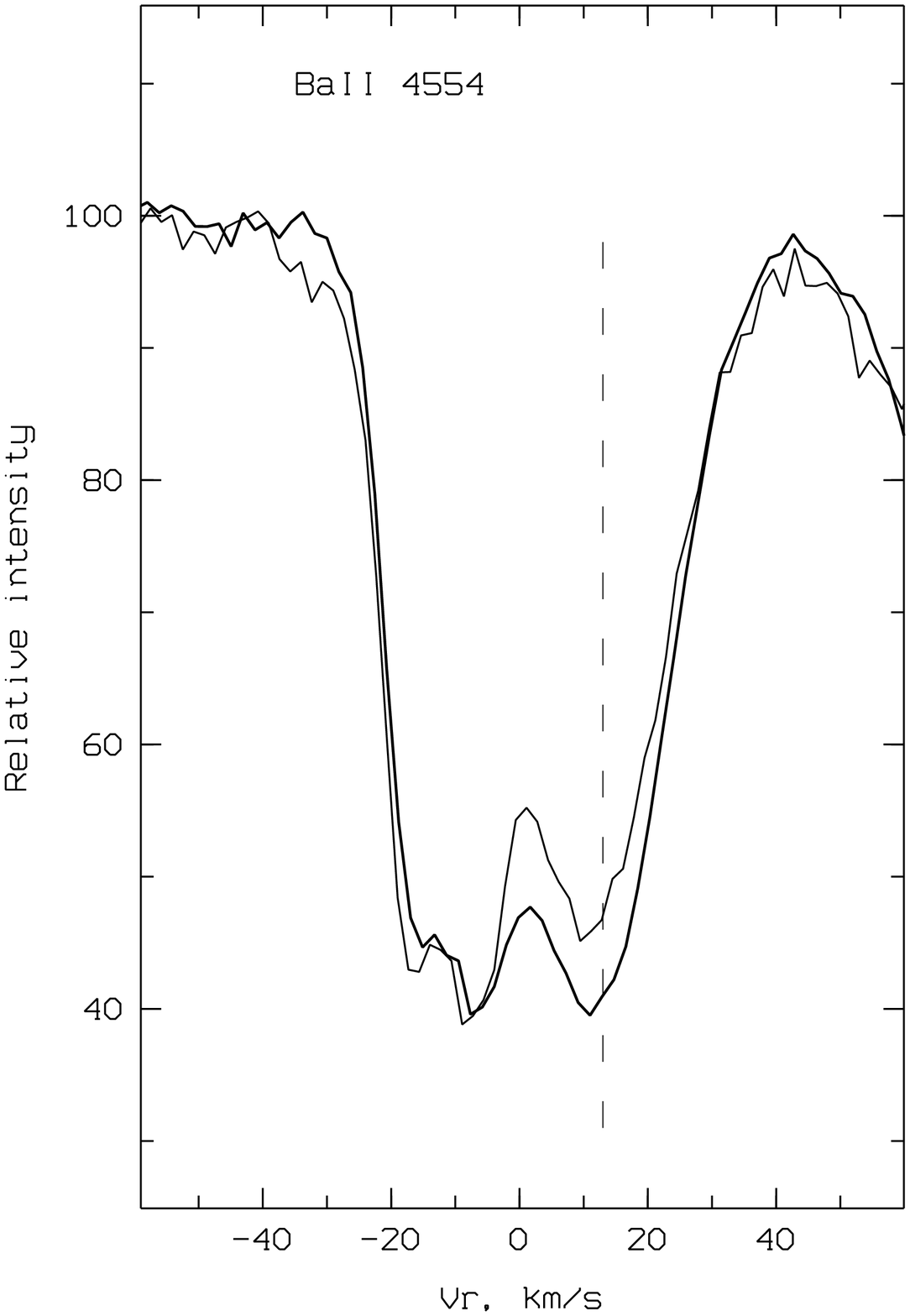}
\includegraphics[angle=0,width=0.4\textwidth,bb=30 50 540 780,clip]{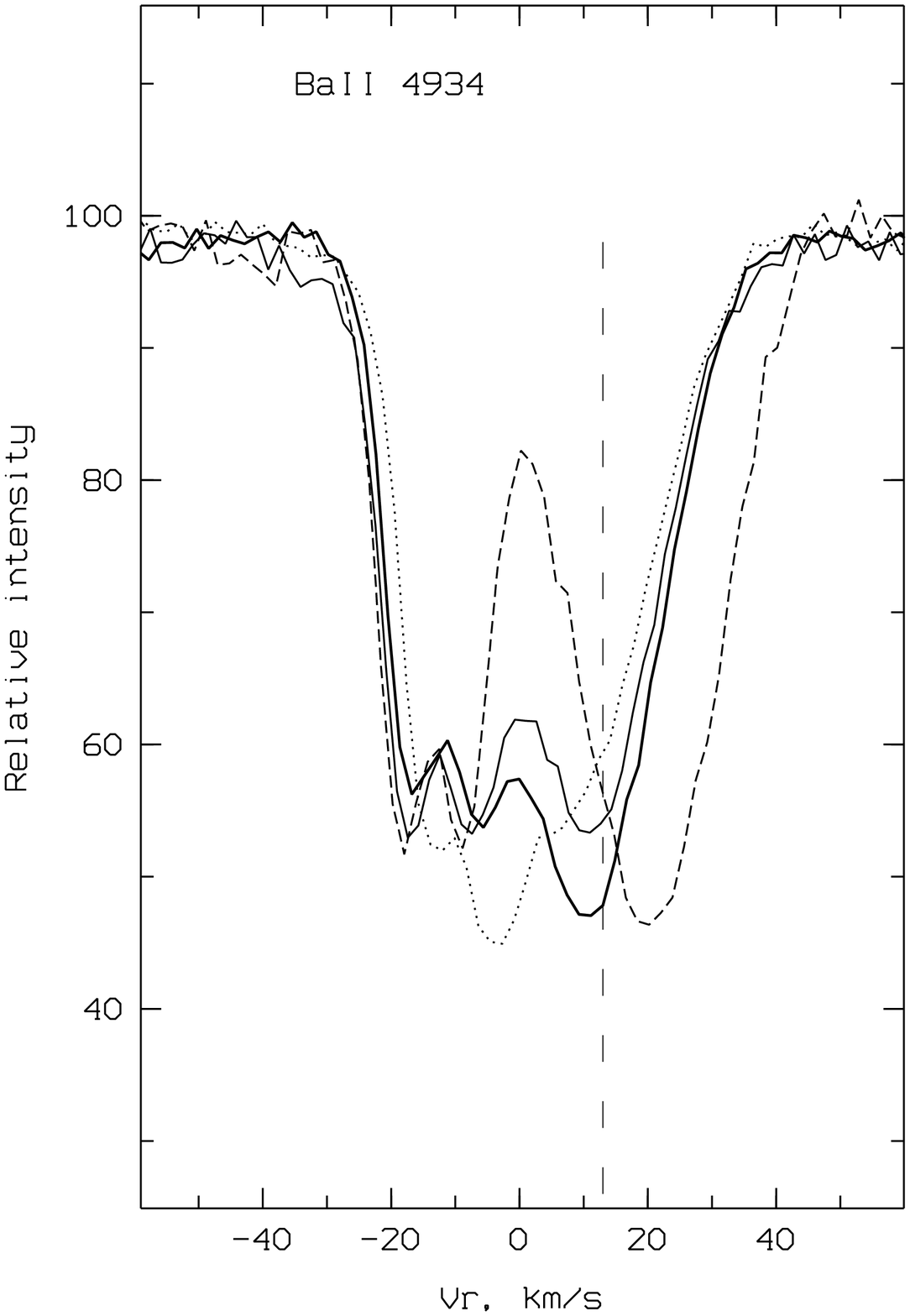}
\caption{Variability of the Ba\,II line profiles in the spectra of V5112\,Sgr taken at different dates: August 2, 2012 
         (thin solid and (b) line), June 13, 2011 (thick solid line), August 14, 2006 (dotted line), and July 7, 2001 
         (dashed line):  left -- Ba\,II\,4554\,\AA{}, right -- the Ba\,II\,4934\,\AA{}.}
\end{figure}

Second, comparison of the line profiles in the spectra of V5112\,Sgr taken on different nights revealed
significant variability of the profile shape and positions of the split--line components. 
To illustrate the variability effect, Fig.\,3 presents the profiles of two Ba\,II
lines with the most prominent profile asymmetry and variability for several dates of observations.

Consider in more detail the picture of radial velocities based on the components of split absorption
lines. Table\,2 presents Vr for the individual components of the Ba\,II lines for four dates of observations.
As follows from our comparison of the data from the last two columns of this table, the positions of
the long--wavelength components of the Ba\,II lines change synchronously with those of the symmetric
absorptions. Thus, we have reason to believe that the red component of the Ba\,II lines
is formed in the stellar atmosphere. The position of the blue component itself changes
only slightly with time, deviating from the mean Vr\,$\approx -16.2$\,km/s  by $\pm0.6$\,km/s. Thus, given the
systemic velocity Vsys\,=\,13\,km/s, we find that the short--wavelength component of the Ba\,II lines (and
other split absorption lines) is formed in layers expanding with a velocity V$_{\rm exp}(1)\approx$\,30\,km/s.

Here, it is important to note that a similar expansion velocity, V$_{\rm exp} \approx$\,30--40\,km/s, was found
by Bujarrabal et al. (1992) based on the recording of the CO emission profiles in the millimeter
spectrum of IRAS\,19500$-$1709 associated with the star V5112\,Sgr. Bujarrabal et al. (1992, 2001)
emphasize that the profiles of the CO bands for IRAS\,19500$-$1709 are two--components ones: a
central narrow emission component originating in a medium with a low (about 10\,km/s) expansion
velocity is superimposed on a high--velocity component. 
In our optical spectra, the split absorption line proﬁles also exhibit a second component located
between the short--wavelength and atmospheric ones: for the Ba\,II lines, the velocity corresponding to the
position of the middle component is, on average, Vr\,$\approx$\,$-7.1$\,km/s. Thus, we obtain the following estimate
for the expansion velocity of the layers in which this component is formed: V$_{\rm exp}(2)\approx$\,20\,km/s.

\begin{figure}  
\includegraphics[angle=0,width=0.45\textwidth,bb=30 50 540 780,clip]{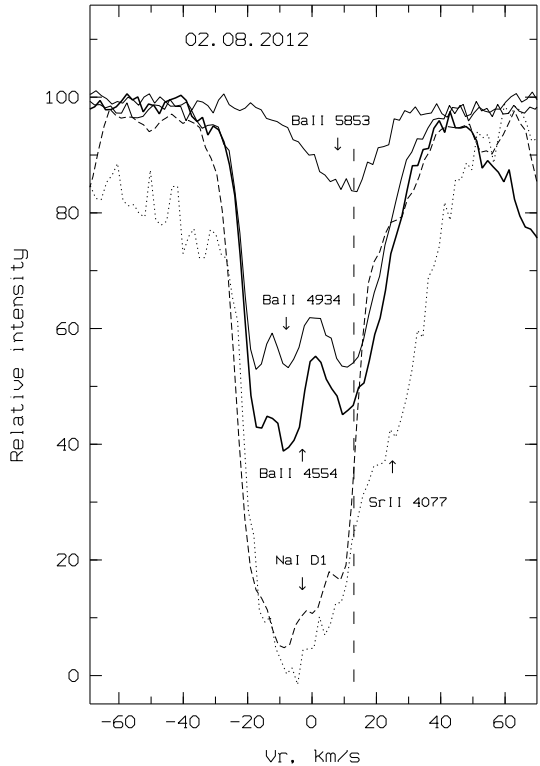}
\caption{Profiles of selected split lines in the August 2, 2012 spectrum of V5112 Sgr.}
\end{figure}

{\bf The Na\,I resonance doublet} was detected for four dates of observations. As follows from Fig.\,5
and the data from columns 5 and 6 of Table\,1, the doublet D line profile is complex. 
Two components are reliably identified at our spectral resolution. The position of the long--wavelength 
component  of the Na\,I~D lines changes with time, being in agreement with the positions of the symmetric 
metal absorptions originating in the stellar atmosphere for each date. In contrast, the position of 
the short--wavelength component of  the Na\,I~D lines is essentially constant with
time. The mean velocity determined from the short--wavelength component is Vr$ -9.0$\,km/s, which
is close to the position of the middle component in the split Ba\,II lines. Obviously, this stable 
short--wavelength component of the Na\,I~D lines is formed in the circumstellar envelope, because we cannot
expect the presence of interstellar Na\,I lines in the spectrum of a star so far away from the Galactic
plane. Bartkevicius (1992) gives the color excess E(B$-$V)\,=\,0.10$^{\rm m}$ in the catalog of UU\,Her stars.
The multiband photometry of V5112\,Sgr performed by Hrivnak et al. (1989) is also indicative of a low
reddening.

\begin{figure}  
\includegraphics[angle=0,width=0.45\textwidth,bb=30 50 540 780,clip]{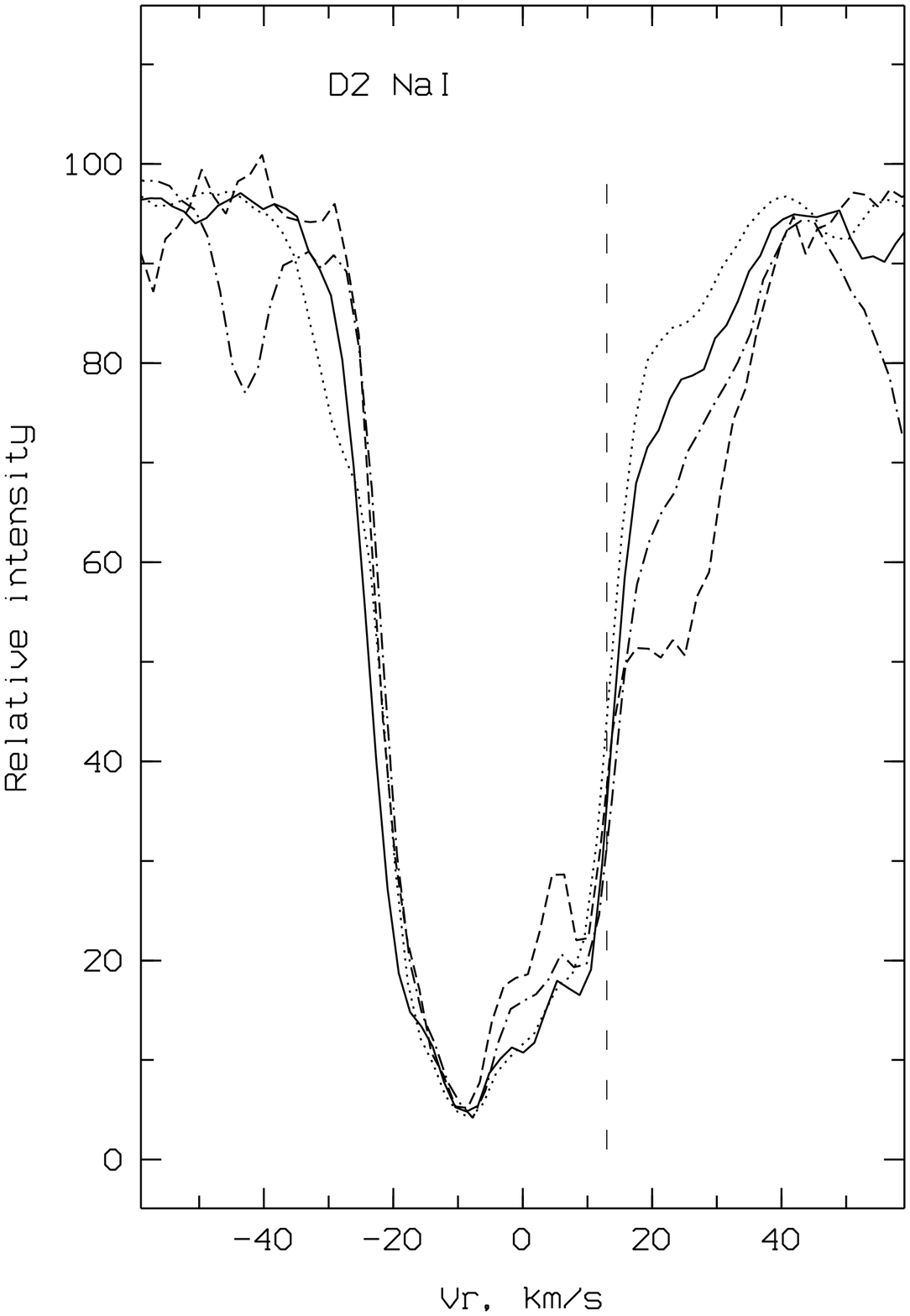}
\includegraphics[angle=0,width=0.45\textwidth,bb=30 50 540 780,clip]{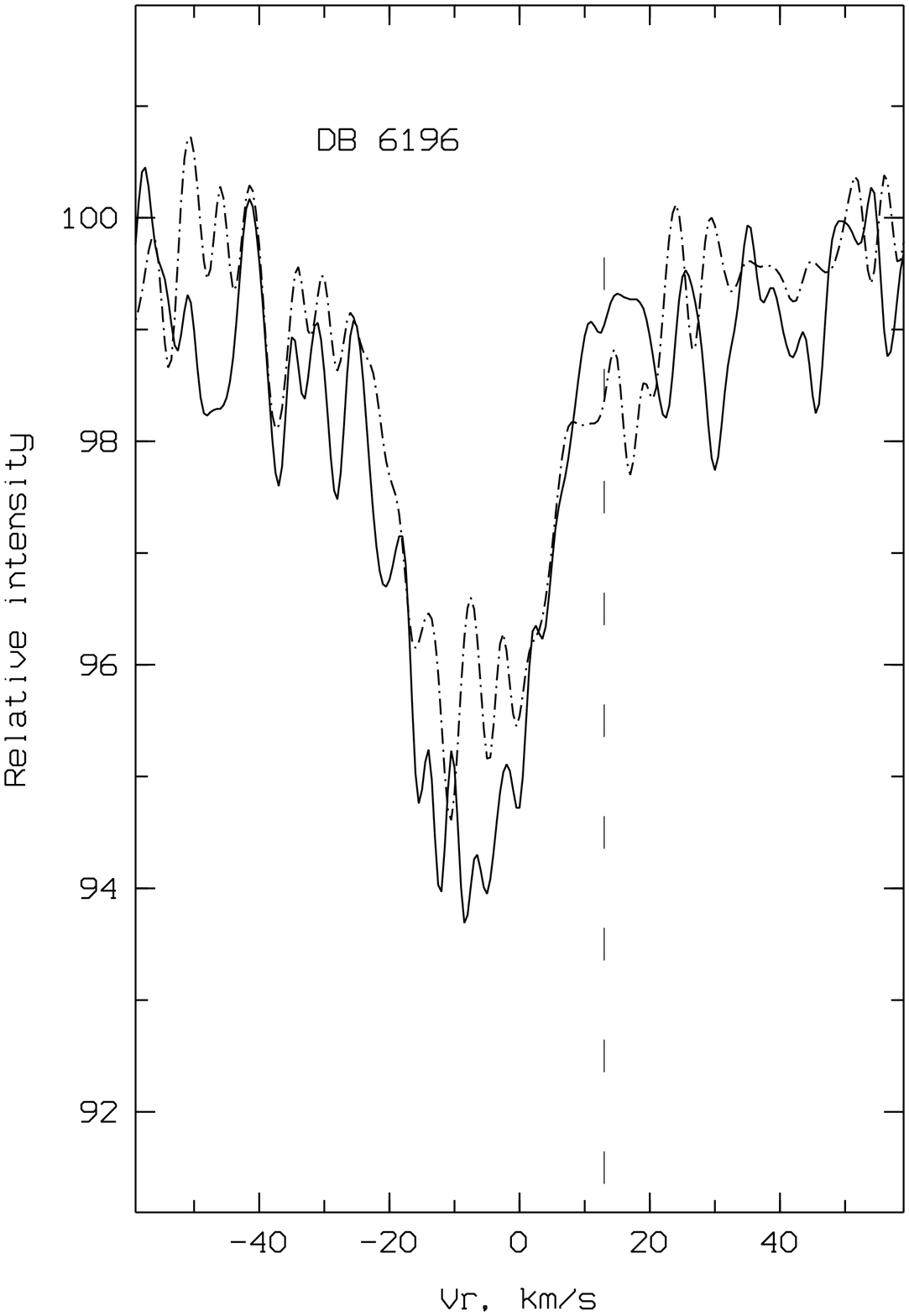}
\caption{Left: profiles of the NaI~D2 5889\,\AA{} line in the spectra of V5112\,Sgr taken at different dates: 
         August 2, 2012 (solid line), September 28, 2010 (dash--dotted line), August 14, 2006 (dotted line), 
         and July 7, 2001 (dashed line) spectra. Right: profile of the diffuse 6195.96\,\AA{} band in the August 2, 2012 
         (solid line) and September 28, 2010 (dash--dotted line) spectra.}
\end{figure}

A complex profile that also contained the circumstellar component, in addition to the photospheric and interstellar ones, 
in the spectra of post--AGB stars with envelopes was previously observed only for the D lines of the Na\,I doublet. 
In particular, this is also true for the star V354\,Lac, in the spectrum of which the circumstellar component of the Na\,I
D lines was identified by Reddy et al. (2002). The profile of the Na\,I D lines and other peculiarities of
the spectrum for this star were studied in more detail by Klochkova (2009) and Klochkova et al. (2009)
based on BTA spectra. The circumstellar absorption components of the Na\,I~D~lines were also identified
in the spectrum of HD\,56126 that was well studied by Bakker et al. (1997), Klochkova et al. (2007b),
and Klochkova and Chentsov (2007). In addition, the cases of a manifestation of the circumstellar 
gaseous--dusty envelope in the form of emission components of the Na\,I~D~lines are known. As an example, we
provide spectra for the star V510\,Pup identified with the infrared source IRAS\,08005$-$2356 (Klochkova
and Chentsov 2004), the bipolar nebula AFGL\,2688 (Klochkova et al. 2004), and the semiregular variable 
QY\,Sge\,=\,IRAS\,20056+1834 (Kameswara Rao et al. 2002; Klochkova et al. 2007c).

We emphasize also that weak absorption features identified with the so--called diffuse bands (DBs) well known 
in the spectroscopy  of the interstellar medium are present in the optical spectrum of V5112 Sgr. 
As an illustration, Fig.\,5 presents the profiles in ``intensity--radial velocity'' coordinates for  
one of the bands at 6195.96\,\AA{} in two spectra. The choice of this band is dictated by the fact that it is 
narrow and least blended in the spectra of F-- and G--supergiants.
Because of the large intensity difference between the  absorption features, the scales of the vertical
axes differ approximately by a factor of 10. This figure suggests that the positions of the circumstellar
component of the Na\,I~D2 line and the 6195.96\,\AA{} band coincide.

In the  the wavelength range 5780--6613\,\AA{}, we measured Vr(DB) from the positions of the reliably 
identified  5780, 5797, 6196, 6234, and 6379\,\AA{} features. We took the exact wavelengths of these bands from
the list by Hobbs et al. (2008). In the August 2, 2012 spectrum,  the equivalent widths of these features are
W\,=\,83, 40, 30, 8, and 18\,м\AA{}, respectively.   The  inaccuracy of the mean velocity determined in the 
spectra of V5112\,Sgr is $\le \pm 0.5$\,km/s. As follows from the last row in Table\,1, 
the mean (over the spectra) Vr(DB)\,=\,$-8.9$\,km/s is in excellent agreement with the velocity derived 
from the short--wavelength component of the Na\,I~D~lines.

The derived coincidence suggests that the weak diffuse bands we detected originate in the circumstellar envelope. 
This fact is not so trivial as it may seem, because after the publication of the results by Luna
et al. (2008), it has been widely believed that the diffuse bands originate only in the interstellar medium;
the physical conditions in the circumstellar envelopes of post--AGB stars are not conducive to their formation.
Luna et al. (2008) reached this conclusion by studying high--resolution spectra for a sample of post--AGB stars. 
We compared our measurements of W and Vr with the data of these authors for V5112\,Sgr
and point out satisfactory agreement in W. However, the measurements of Vr for V5112\,Sgr from Luna
et al. (2008) have a low accuracy due to the inclusion of the wide band at 6284\,\AA{}, which, 
besides, is blended with telluric features, and the use of the band  at 6613.7\,\AA{}, which is located 
in the blue wing of a very strong Y\,II absorption line in the spectrum of V5112\,Sgr. Yttrium has a large 
overabundance in the stellar atmosphere: according to van Winckel (1997),
[s/Fe]\,=\,+1.3. Therefore, the 6613.7\,\AA{} line profile has anomalies typical of the strong absorption 
lines of heavy metals in the spectrum of V5112 Sgr.

\section{DISCUSSION}

As we have pointed out above, an optical manifestation of circumstellar envelopes has been found
previously in the D lines of the Na\,I doublet. The results of observations of the circumstellar envelopes 
around Mira--type stars in the K\,I 7665 and 7669\,\AA{} lines are also known (see Guilain and Mauron (1996)
and references therein). Based on high--resolution spectra, apart from Na\,I and K\,I, Mauron and Huggins (2010) 
also detected other metals in the gas phase in the Mira--type star CW\,Leo (the infrared source IRC\,+10216), 
which has an extremely extended structured envelope with an angular size of about 200\,arcsec: Ca\,I, Ca\,II, and Fe\,I.

\begin{table}
\bigskip
\caption{Basic information for four related post-AGB stars }    
\bigskip
\begin{tabular}{ l| c|   c| l|  l| l }          
\hline
 Star   & 21\,$\mu$& [C/Fe] и & Envelopes & \multicolumn{2}{c}{$V_{\rm exp}$, km/s}\\ 
\cline{5-6}
           &        &  [s/Fe] & structure  &  \hspace{1cm} CO  & optical spectra   \\
\hline                                                                   
  \hline
CY\,CMi    &   +    & +   &halo         & 10--12  (L1993)      & 11 (K2007)  \\
V5112\,Sgr &   +    & +   &bipolar     & 8, 11--14.4 (L1993)  & 20 и 30 (this paper)\\
           &        &     &               & 10, 30--40 (B1992)   &   \\
V354\,Lac  &   +    & +   &bipolar     & 9.6--11.6 (L1993)    & 10.8 (K2009)  \\
V448\,Lac  &   +    & +   &halo\,+\,arcs  & 14--15 (L1993)       & 15.2 (K2010)  \\ 
\hline
\multicolumn{6}{l}{} \\
\multicolumn{6}{l}{\footnotesize\it   Note to the table. The overabundances of carbon [C/Fe] and heavy metals [s/Fe] are marked by “+” in column 3 }\\
\multicolumn{6}{l}{\footnotesize\it   in accordance with the results from Klochkova (1995) for CY CMi, van Winckel et al. (1996)}\\
\multicolumn{6}{l}{\footnotesize\it   for V5112 Sgr,  Klochkova et al. (2009) for V354 Lac, and Decin et al. (1998) for V448 Lac. }\\
\multicolumn{6}{l}{\footnotesize\it   The type of circumstellar envelope morphology  (column 4) is given according to the Hubble space}\\
\multicolumn{6}{l}{\footnotesize\it   telescope observations by Sahai et al. (2007). The expansion velocities V$_{\rm exp}$ in column 5 are given}\\
\multicolumn{6}{l}{\footnotesize\it   based on CO band observations.  The last column gives the expansion velocity determined from the}\\
\multicolumn{6}{l}{\footnotesize\it   positions of the Swan C$_2$ molecular bands in the optical spectra of CY\,CMi,  V354 Lac, and V448 Lac }\\
\multicolumn{6}{l}{\footnotesize\it   and from the shell  components of the Ba II lines for V5112 Sgr.  The references are designated as L1993  }\\
\multicolumn{6}{l}{\footnotesize\it   for Loup et al. (1993), B1992 for Bujarrabal et al. (1992),  K2007 for Klochkova et al. (2007b), }\\
\multicolumn{6}{l}{\footnotesize\it   K2009 for Klochkova et al. (2009), and K2010 for Klochkova et al. (2010).}\\
\end{tabular}
\end{table}

In this work, we have detected a manifestation of the envelope in lines of heavy metals in the
spectrum of a star at a later evolutionary stage for the first time. The main new result of our
study is the detection of a multicomponent Ba\,II line profile and an asymmetry of the low--excitation
absorption lines for a number of other metals in the spectrum of V5112\,Sgr. Previously, we found a
distortion of the profiles of the strongest absorption lines for three more post--AGB stars with similar 
parameters: CY\,CMi\,=\,IRAS\,07134+1005, V354\,Lac\,=\,22274+5435, and 
V448\,Lac\,=\,22223+4327.

Table 3 gives basic properties for these related post--AGB stars. In addition to the common signatures of 
protoplanetary nebulae (PPN) formulated by Kwok (1993), a further study of these stars and
associated infrared sources revealed a number of important properties. In particular, an emission 
feature at $\lambda = 21\mu$ was detected in the infrared spectra of all four objects being compared. 
The 21\,$\mu$ band was first detected in the infrared spectra of four PPNs by Kwok et al. (1989); a 
decade later, the same authors (Kwok et al. 1999) published a list of already a dozen
PPN, among which there are also the four stars from Table\,3.  Based on high--spatial--resolution Hubble
space telescope snapshots, Sahai et al. (2007) and Ueta et al. (2000, 2008) concluded that the envelopes
of these four stars have different aspherical morphologies (see column 4 in Table 3). The last two columns
in this table give the expansion velocity V$_{\rm exp}$ of the circumstellar envelope from millimeter (column 5)
and optical (column 6) observations.

Among the objects listed in Table\,3, the semiregular variable V354\,Lac (spectral type G5\,Iap, as
inferred by Hrivnak (1995)) is closely analogous to V5112 Sgr in envelope structure and spectrum peculiarities. 
The spectroscopic monitoring of this star performed by Klochkova et al. (2009) at the 6-m
BTA telescope with resolution R\,=\,60000 revealed a splitting of the strongest absorption lines with a
lower--level excitation potential $\chi_{\rm low}< 1$\,eV. Analysis of the kinematics showed that the 
short--wavelength component of the split line is formed in the star extended gas--dust envelope. 
This splitting is most  clearly seen in the profile of the Ba\,II~6141\,\AA{} line with an equivalent width 
W$\approx$1\,\AA{}. The Doppler separation between the absorption components of this line is about 35\,km/s. 
The position of the short--wavelength component of Ba\,II coincides with the position of the circumstellar 
component in the profile of the Na\,I D-lines, which is formed in the same layers where the Swan C$_2$ molecular 
bands are also formed. This coincidence suggests that, apart from the atmospheric component, 
the complex Ba\,II\,6141\,\AA{} line profile contains the component originating in the circumstellar envelope. 
Such a splitting (or asymmetry of the profile because of the flatter blue wing) is also observed for other 
Ba\,II lines  ($\lambda$\,4554, 5853, and 6496 A) as well as for the strong  La\,II\,6390\,\AA{},
Nd\,II\,5234\,\AA{}, 5293\,\AA{},  Y\,II\,5402\,\AA{}, lines. In the spectrum of V354\,Lac, the lines of 
these heavy--element ions are enhanced to an extent that their intensities are comparable to those of the neutral
hydrogen lines.

Based on BTA spectra, Klochkova et al. (2007b) and Klochkova et al. (2010) found peculiarities and
variability of the optical spectra for two other stars from Table\,3, CY\,CMi and V448\,Lac, respectively.
The profiles of the strongest metal absorption lines in the spectrum of these two stars are asymmetric. All
lines, except the hydrogen ones, have the same type of asymmetry: the blue wing is more extended than
the red one, i.e., the radial velocity measured from the upper part of the profile (from the wings) turns out
to be lower than that measured from the core. The blueshift of the wings relative to the core increases
as the line strengthens, gradually or abruptly. In addition, we see from our comparison of the spectra
taken at different dates that the profile shape of these lines originating in the star expanding atmosphere
(at the base of the wind) changes both with time and with line intensity. We emphasize that no clear splitting 
into components has been found in the collection of BTA echelle spectra for CY\,CMi and V448\,Lac
even for the most intense absorption lines. The expansion velocity was measured with a high accuracy
for both stars from several tens of rotational lines of the Swan C$_2$ molecular bands and it was concluded
that the molecular envelope expansion is stable. 

Note that a probable splitting of the absorption lines into components with different positions and intensities 
for different dates of observations must not be ruled out a priori. Observations with an ultra--high spectral 
resolution are needed to test this assumption.

The difference in profile peculiarity type for two pairs of stars, V5112\,Sgr and V354\,Lac with split
profiles of the strongest absorption lines of selected elements as well as CY\,CMi and V448\,Lac with
asymmetric but unsplit profiles, suggests that a complex morphology of the circumstellar envelope could be
a factor causing the profile peculiarity and variability for strong lines. We see from the data in Table~3 that
the envelope for the pair of V5112\,Sgr and V354\,Lac with split absorption lines is bipolar, while for CY\,CMi
and V448\,Lac there is no splitting and the extended envelopes have a less distinct structure. The validity
of this assumption is also confirmed by the fact that in the spectrum of V5112\,Sgr, for which Bujarrabal et al. 
(1992) detected slow (V$_{\rm exp}$\,=\,10\,km/s) and fast (30--40\,km/s) expansion based on CO bands, we see a three--component 
profile structure for strong absorption lines. The split line profiles include the photospheric and two shell 
components, one of which, by analogy with the CO profile, originates in the envelope formed at the AGB stage and expanding
with a velocity V$_{\rm exp}(2) \approx$\,20\,km/s and the other originates in the envelope with an expansion velocity
V$_{\rm exp}(1)\approx$\,30\,km/s formed later.

At present, there are no universally accepted views of the formation of deviations from spherical
symmetry for PPN. An asymmetry (in particular, bipolarity) of the structures in selected PPN was
detected in several types of observations. Having investigated the polarization of the near--infrared
emission for a sample of extended nebulae, Gledhill et al. (2001) identified three different shapes of these
objects: isotropic, bipolar, and nebulae with a dominant core. IRAS\,19500$-$1709 in polarized light is
a typical bipolar nebula with two lobes and a dust lane. Having analyzed high--spatial--resolution HST
optical images for a sample of PPN, Ueta et al. (2000, 2008) concluded that the optical thickness of 
the circumstellar matter is a decisive factor in the formation of a particular type of stellar envelope morphology.

A brief overview of the physical mechanisms for the formation of these complex structures can be
found, for example, in Lagadec et al. (2011). As a rule, it is assumed that the dense spherical envelope
formed at the AGB stage expands with a low velocity, while the axisymmetric part of the envelope formed
later, at the post--AGB stage, undergoes fast expansion. The sequence of these processes gives rise to
a gradient of the optical thickness in the direction from the equator to the polar axis of the system. The
existence of a companion in the system and/or the presence of a magnetic field can also be the physical
factors due to which the spherical symmetry of the stellar envelope is lost in the short interval of evolution
between AGB and post--AGB (see Huggins et al. 2009; Leal--Ferreira et al. 2012; and references
therein). Recently, Koning et al. (2013) proposed a simple PPN model that was constructed based on a
pair of evacuated cavities inside a dense spherical halo. The authors showed that all of the morphological
peculiarities observed in real bipolar PPNs could be reproduced by varying parameters (the matter density
in the cavity, its size and orientation). 

As Sahai et al. (2007) pointed out, the difference in the shape of ``extended halo'' and ``bipolar'' nebulae 
could be purely visual, because the observed shape depends strongly on the inclination of the axis to the line 
of sight and on the angular resolution of the structure. For instance, according to Nakashima et al. (2009), 
the extended envelope of CY\,CMi has no clear structure in the form of jets, cavities, etc.

In future, it is especially important to study with a high resolution (R\,$\ge$\,60000) the optical spectra
of other related stars from the sample of objects with the 21\,$\mu$ emission feature with atmospheres
enriched with heavy s--process metals. In particular, this also applies to the faint stars with carbon enriched 
envelopes identified with the infrared sources IRAS\,04296+3429 and 23304+6147. According to the data by Klochkova 
et al. (1999, 2000), the atmospheres of the central stars of both sources are enriched with carbon and heavy metals, 
while according to the observations by Sahai et al. (2007), the envelope of IRAS\,04296+3429 has a bipolar shape
and the envelope of IRAS 23304+6147 probably has a quadrupole shape.

\section{Conclusions}

Based on our high--resolution echelle spectra of the post--AGB supergiant V5112\,Sgr taken in 
1996--2012 at the 6--m BTA telescope, we detected peculiarities and variability of the spectrum and the velocity
field in the stellar atmosphere and envelope. An asymmetry and splitting of strong metal absorption
lines, which is maximal for the Ba\,II lines whose profile is split into three components, have been found
for this star for the first time. The profile shape of the split lines and their positions are variable in time.

We determined two envelope expansion velocities V$_{\rm exp} \approx$\,20 and 30\,km/s consistent with the picture 
of envelope  expansion from millimeter observations. The coincidence of Vr determined from the short--wavelength 
components of the split  absorption lines with the velocities derived from CO molecular bands and the 
short--wavelength component of the Na I D lines suggests that the short--wavelength components of the heavy--metal 
lines are formed in a structured circumstellar envelope. Thus, evidence for an efficient dredge--up of the heavy
metals produced during the star’s preceding evolution into the envelope has been obtained for the first time.

Analysis of the set of radial velocities Vr inferred from symmetric absorption lines confirmed the presence of 
pulsations in the stellar atmosphere with a pulsation amplitude $\Delta$Vr$\le 8$\,km/s.

The mean (from three spectra) radial velocity determined from diffuse bands in the spectrum of
V5112\,Sgr agrees excellently with the velocity derived from the short--wavelength shell component of
the Na\,I~D lines. This leads us to conclude that the diffuse bands are formed in the circumstellar
envelope.

\section*{Acknowledgements}

This work was supported by the Russian Foundation for Basic Research (project 11--02--00319\,a)
and the ``Extended Objects in the Universe'' Basic Research Program of the Division of Physical 
Sciences of the Russian Academy of Sciences.

\newpage

\centerline{\Large\bf References}

\begin{itemize}

\item{} V.P. Arhipova, N.P. Ikonnikova, and G.V. Komissarova, Astron. Lett. {\bf 36} 268 (2010).

\item{}   E.J. Bakker, E.F. van Dishoeck, L.B.F.M. Waters, and T. Schoenmaker, Astron. Astrophys. {\bf 323} 469
     (1997).

\item{}   A. Bartkevicius, Baltic Astron. {\bf 1} 194 (1992).

\item{}  V. Bujarrabal, J. Alcolea, and P. Planesas, Astron. Astrophys. {\bf 257} 701 (1992).

\item{}   V. Bujarrabal, A. Castro-Carrizo, J. Alcolea, and C. Sanchez Contreras, Astron. Astrophys. {\bf 377} 868
(2001).

\item{}  L. Decin, H. van Winckel, C. Waelkens, and E. J. Bakker, Astron. Astrophys. {\bf 332} 928 (1998).

\item{}  G.A. Galazutdinov, Preprint Spec. Astrophys. Observ. No.\,92 (1992).

 \item{} T.M. Gledhill, A. Chrysostomou, J.H. Hough, and J.A. Yates, Mon. Not. R. Astron. Soc. {\bf 322} 321
(2001).

\item{}  Ch. Guilain and N. Mauron, Astron. Astrophys. {\bf 314} 585 (1996).

\item{}  L.M. Hobbs, D.G. York, T.P. Snow, et al., Astrophys. J. {\bf 680} 1256 (2008).

\item{}  B.J. Hrivnak, Astrophys. J. {\bf 438} 341 (1995).

\item{}  B. Hrivnak, S. Kwok, and K. M. Volk, Astrophys. J. {\bf 346} 265 (1989).

 \item{} B.J. Hrivnak, W. Lu, R.E. Maupin, and B.D. Spitzbart, Astrophys. J. {\bf 709} 1042 (2010).

\item{}  B.J. Hrivnak, W. Lu, K.L. Wefel, et al., Astrophys. J. {\bf 734} 25 (2011).

 \item{} P.J. Huggins, N. Mauron, and E.A. Wirth, Mon. Not. R. Astron. Soc. {\bf 396} 1805 (2009).

\item{}  N. Kameswara Rao, A. Goswami, and D.L. Lambert, Mon. Not. R. Astron. Soc. {\bf 334} 129 (2002).

\item{}  V.G. Klochkova, Mon. Not. R. Astron. Soc. {\bf 272} 710 (1995).

\item{}  V.G. Klochkova, Bull. Spec. Astrophys. Obs. {\bf 44} 5 (1997).

 \item{}  V.G. Klochkova, Astron. Lett. {\bf 35} 457 (2009).

\item{}  V.G. Klochkova, Astrophys. Bull.  {\bf  67} 385 (2012).

\item{}  V.G. Klochkova and V.E. Panchuk, Astron. Rep. {\bf 56} 104 (2012).

\item{}  V.G. Klochkova and N.S. Tavolganskaya, Astrophys. Bull. {\bf 65} 18 (2010).

\item{}  V.G. Klochkova and E.L. Chentsov, Astron. Rep. {\bf 48} 301 (2004).

\item{}  V.G. Klochkova and E.L. Chentsov, Astron. Rep. {\bf 51} 994 (2007).

\item{}  V.G. Klochkova, R. Szczerba, V.E. Panchuk, and K. Volk, Astron. Astrophys. {\bf 345} 905 (1999).

\item{} V.G. Klochkova, R. Szczerba, and V.E. Panchuk, Astron. Lett. {\bf 26} 88 (2000).

\item{}  V.G. Klochkova, M.V. Yushkin, A.S. Miroshnichenko, V.E. Panchuk, and K. Bjorkman, Astron.
Astrophys. {\bf 382} 143 (2002).

\item{} V.G. Klochkova, V.E. Panchuk, M.V. Yushkin, and A.S. Miroshnichenko, Astron. Rep. {\bf 48} 288 (2004).

\item{} V.G. Klochkova, V.E. Panchuk, N.S. Tavolzhanskaya, and G. Zhao, Astron. Rep. {\bf 50} 232 (2006).

\item{} V.G. Klochkova, E.L. Chentsov, N.S. Tavolzhanskaya, and V.E. Panchuk, Astron. Rep. {\bf 48} 642 (2004a).

\item{}  V.G. Klochkova, E.L. Chentsov, N.S. Tavolganskaya, and M.V. Shapovalov, Astrophys.
    Bull. {\bf 62} 162 (2007b).

\item{}  V.G. Klochkova, V.E. Panchuk, E.L. Chentsov, and M.V. Yushkin, Astrophys. Bull.{\bf 62} 217
(2007c).

\item{}  V.G. Klochkova, E.L. Chentsov, and V.E. Panchuk, Astrophys. Bull. {\bf 63} 112 (2008a).

\item{}  V.G. Klochkova, V.E. Panchuk, M.V. Yushkin, and D.S. Nasonov, Astrophys. Bull. {\bf 63} 386
        (2008b).

\item{}  V.G. Klochkova, V.E. Panchuk, and N.S. Tavolganskaya, Astrophys. Bull. {\bf 64} 155 (2009).

\item{}  V.G. Klochkova, V.E. Panchuk, and N.S. Tavolzhanskaya, Astron. Rep. {\bf 54} 234 (2010).

\item{}  N. Koning, S. Kwok, and W. Steffen, Astrophys. J. {\bf 765} 92 (2013).

\item{}  S. Kwok, Ann. Rev. Astron. Astrophys. {\bf 31} 63 (1993).

\item{} S. Kwok, K.M. Volk, and B. Hrivnak, Astrophys. J. {\bf 345} L51 (1989).

\item{}  S. Kwok, K.M. Volk, and B. Hrivnak, in Asymptotic Giant Branch Stars, Proceedings of the IAU
Symposium No. 191, Ed. by T. Le Bertre, A. Lebre, and C. Waelkens (Astron. Soc. of the Paciﬁc, San
Francisco, USA, 1999), p. 297.

\item{}  E. Lagadec, T. Verhoelst, D. Mekarnia, et al., Mon. Not. R. Astron. Soc. {\bf 417} 32 (2011).

\item{} M.L. Leal-Ferreira, W.H.T. Vlemmings, P.J. Diamond, et al., Astron. Astrophys. {\bf 540} A42 (2012).

\item{} C. Loup, T. Forveille, A. Omont, and J.P. Paul, Astron. Astrophys. Suppl. Ser. {\bf 99} 291 (1993).

\item{} R. Luna, T. L.X. Cox, M.A. Satorre, et al., Astron. Astrophys. {\bf 480} 133 (2008).

\item{}   N. Mauron and P.J. Huggins, Astron. Astrophys. {\bf 513} A31 (2010).

\item{} J. Nakashima, N. Koning, S. Kwok, and Y. Zhang, Astrophys. J. {\bf 692} 402 (2009).

\item{}  V.E. Panchuk, V.G. Klochkova, G.A. Galazutdinov, V.P. Ryadchenko, and E.L. Chentsov, Astron. Lett.
    {\bf 19} 431 (1993).

\item{} V. Panchuk, V. Klochkova, M. Yushkin, and I. Najdenov, in The UV Universe: Stars from Birth to
Death, Proceedings of the Joint Discussion No. 4 during the IAU General Assembly of 2006, Ed. by
A. I. Gomez de Castro and M. A. Barstow (Editorial Complutense, Madrid, 2007), p. 179.

\item{}  V.E. Panchuk, V.G. Klochkova, M.V. Yushkin, and I.D. Naidenov, J. Optical Technology, {\bf 76} 42 (2009).

\item{}  M. Parthasarathy, S.R. Pottash, and W. Wamsteker, Astron. Astrophys.{\bf  203} 117 (1988).

\item{}  B.E. Reddy, M. Parthasarathy, G. Gonzalez, and E.J. Bakker, Astron. Astrophys. {\bf 328} 331 (1997).

\item{} B.E. Reddy, D.L. Lambert, G. Gonzalez, and D. Yong, Astrophys. J. {\bf 564} 482 (2002).

\item{}  R. Sahai, M. Morris, C. Sanchez Contreras, and M. Glaussen, Astron. J. {\bf 134} 2200 (2007).

\item{}  C. Sanchez Contreras, R. Sahai, A. Gil de Paz, and R. Goodrich, Astrophys. J. Suppl. Ser. {\bf  179} 166
          (2008).

\item{}  D.D. Sasselov, Astrophys. Space Sci. {\bf 102} 161 (1984).

\item{}  N. Siodmiak, M. Meixner, T. Ueta, et al., Astrophys. J. {\bf 677} 382 (2008).

\item{}  T. Ueta, M. Meixner, and M. Bobrowsky, Astrophys. J. {\bf 528} 861 (2000).

\item{}  W.E.C.J. van der Veen, N.R. Trams, and L.B.F.M. Waters, Astron. Astrophys.  {\bf 269} 231
           (1993).

\item{}  H. van Winckel, Astron. Astrophys. {\bf 319} 561 (1997).

\item{} H. van Winckel, C. Waelkens, and L.B.F.M. Waters, Astron. Astrophys. {\bf 306} L37 (1996).

\item{} M.V. Yushkin and V.G. Klochkova, Preprint Spec. Astrophys. Observ. No.\,206 (2005).

\end{itemize}

\end{document}